Article

# Toward a Unified Theory of Catalysis

Frank N. Crespilho[1*]

Catalysis lies at the heart of chemical reactivity, yet its foundational principles remain fragmented across the distinct domains of homogeneous, heterogeneous, and enzymatic systems[1-8]. Here, we propose a unifying theoretical model that integrates spatial and temporal dimensions into a single framework, offering a cohesive understanding of catalytic activity across diverse materials and conditions. This model builds upon established kinetic theories, incorporating local site density distributions, time-dependent modulations, and intrinsic reaction rates to deliver a comprehensive description of catalytic performance. By applying this approach, we demonstrate how seemingly disparate catalytic processes—from molecular complexes and single-atom catalysts to complex enzyme systems—can be interpreted through shared physical and chemical principles.

Catalysis has evolved through successive breakthroughs, from Döbereiner's hydrogen ignition and Berzelius's coining of the term "catalysis" in the 19th century to the industrial-scale ammonia synthesis of the Haber–Bosch process and catalytic cracking in the early 20th century. Enzymatic catalysis advanced mid-century with the Michaelis–Menten model and metalloenzyme studies, while the late 20th century brought major developments in homogeneous catalysis, including proline-catalyzed aldol reactions[1] and asymmetric alkylation via photoredox-organocatalysis[2]. In parallel, directed enzyme evolution[3], self-assembled monolayers[4], and supramolecular approaches[5] expanded the mechanistic repertoire and structural complexity of catalysts. More recently, dynamic structural evolution in electrocatalysts[6] and adaptive homogeneous cross-coupling systems[7] have deepened our understanding of catalytic adaptability under operating conditions. These innovations build upon foundational insights from early studies in heterogeneous catalysis[8], laying the groundwork for cross-disciplinary integration.

Over time, enzymology, homogeneous catalysis, and heterogeneous catalysis developed specialized principles—such as active-site conformational changes, ligand-field effects, and surface adsorption thermodynamics—yet these fields often lack a unifying conceptual framework, limiting the transfer of insights and design rules across disciplines. However, experimental advances now paint a different picture—one where reactivity is spatially extended and temporally dynamic. Enzymes exhibit allosteric regulation and proton-coupled electron transfer through evolving tertiary structures. Homogeneous catalysts form fluctuating ensembles responsive to solvation and concentration gradients. Solid-state systems reveal localized "hot spots", restructuring under *operando* conditions, and electronic coupling with support that modulate activity at the nanoscale. Such evidence supports a shift in perspective: catalysis is better understood not as a discrete pointwise interaction, but as an emergent property of a spatially continuous and temporally evolving field. This view aligns with observations from modern o*perando* spectroscopy, single-molecule microscopy, and electrochemical imaging, which consistently reveal heterogeneous patterns of reactivity within seemingly uniform systems. The *active catalytic space* framework[9] embraces this complexity, offering a generalizable and quantitative description of catalysis that spans biological, molecular, and materials chemistry.

By formalizing this perspective through the theory of the active catalytic space, we reframe catalysis as a volumetric phenomenon governed by three interdependent elements: (i) *Site Distribution ($\rho(r)$):* the spatial density of catalytically competent units; (ii) *Dynamic Modulation ($f(r, t)$)*: the accessibility or activity of these sites as modulated by structural, electronic, or environmental factors; (iii) *Local Kinetics ($k_{local}(r, t)$):* the intrinsic rate at which individual sites convert substrate to product. Together, these components define a time-dependent field of catalytic activity, allowing the calculation of integrated metrics such as TOF and TON via spatial and temporal integration (Fig. 1).

This framework provides a basis for comparing catalytic systems that previously appeared unrelated. A solid-state catalyst with nanoscopic hot spots can be described with high $\rho(\vec{r})$ in select regions and variable $k_{local}$, while an enzyme with slow conformational gating may exhibit uniform $\rho(\vec{r})$ but a highly structured and temporally gated $f(\vec{r}, t)$. Both are instances of the same underlying phenomenon: catalysis as a spatiotemporal orchestration of structure and reactivity. We demonstrate this implementation across a diverse range of catalytic systems, including enzymatic catalysts such as carbonic anhydrase and lactate dehydrogenase, homogeneous molecular complexes in solution, atomically dispersed electrocatalysts, and photoactive perovskite-based electrodes under operando conditions. All systems were analyzed under both steady-state and time-dependent regimes to extract spatial and temporal resolution.

[1] São Carlos Institute of Chemistry, University of São Paulo (USP), São Carlos, SP, 13566-590 Brazil

# Article

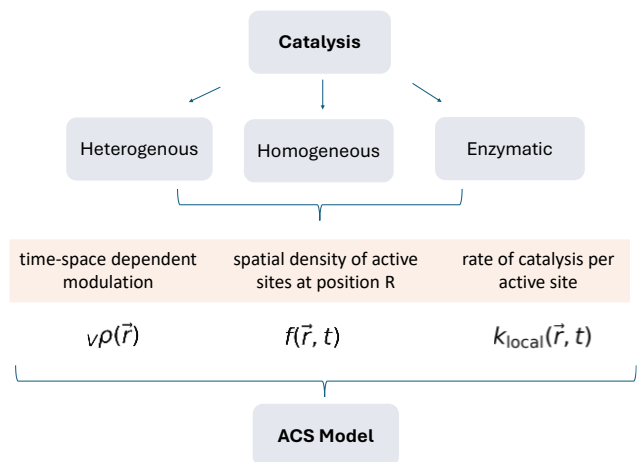

**Figure 1. Integration of Catalytic Domains into the Active Catalytic Space (ACS) Framework.** Schematic representation of the conceptual unification of heterogeneous, homogeneous, and enzymatic catalysis through the ACS model. Each catalytic class is characterized by distinct spatial, structural, and kinetic features, yet all can be described by the generalized volumetric expression of TOF. This framework allows catalytic activity to be treated as a spatially and temporally distributed field, providing a unified basis for analyzing and comparing complex catalytic systems.

## Theory and Mathematical Model

A major conceptual advance in the definition of catalytic efficiency was proposed by Kozuch and Martin,[10] who revisited the foundational concepts of turnover frequency (TOF) and turnover number (TON) across enzymatic, homogeneous, and heterogeneous catalysis. In their view, TOF should not be treated merely as a ratio of product formation over time, but rather as a differential quantity—the derivative of the number of catalytic turnovers with respect to time, under well-defined conditions. To allow cross-system comparison, they introduced the standard turnover frequency (TOF°), measured at 1 M concentration of reactants and products, at 273.15 K, and under steady-state conditions with a fully active catalyst. The proposed expression for TOF° is derived from transition state theory:

$$\text{TOF}° = \frac{k_B T}{h} \cdot e^{-\delta E / RT} \quad (1)$$

where δE is the energetic span of the catalytic cycle. This formulation allows catalytic activity to be expressed in a way analogous to standard thermodynamic functions such as ΔG°, enabling more rigorous benchmarking of catalysts based on intrinsic performance rather than experimental conditions.

While the TOF° is powerful in its simplicity and utility for benchmarking catalysts, it ultimately reduces catalytic behavior to a single scalar parameter—the energetic span between key intermediates and transition states. This formulation implicitly assumes a dominant reaction pathway, ideal and homogeneous reaction conditions, a representative catalytic center, and no explicit spatial or temporal variation. While adequate for standardized comparisons, such an approach overlooks the complex, dynamic nature of real catalytic systems. As acknowledged by Kozuch and Martin—*'this standard TOF is not meant to be a universal archetype of catalytic efficiency, immune to any specific circumstance of the measurement'*—a single TOF value cannot capture the full complexity of catalytic behavior across nonlinear regimes, thus motivating the need for a more general framework. A true generalization of catalytic activity should offer a more comprehensive and physically grounded description—one that reflects the intrinsic heterogeneity, temporal modulation, and spatial organization found in many catalytic environments. A full derivation is provided in the **SI**, where we show that the standard TOF° expression by Kozuch and Martin can be rigorously derived as a limiting case of the ACS formalism. This is achieved by assuming spatial homogeneity, full catalyst activation, and a single energetic span dominating the catalytic cycle.

Thus, we define the active catalytic space as a three-dimensional (3D) volume, $V \subset \mathbb{R}^3$, where catalysis occurs. Within this volume, we describe the following fields:

I. Catalytic site density: $\rho(\vec{r})$ – the spatial density of active sites at position R
II. Dynamic modulation function: $f(\vec{r}, t)$ – a time- and space-dependent modulation of site accessibility, conformation, or activity.
III. Local reaction rate (per site): $k_{\text{local}}(\vec{r}, t)$ – the rate of catalysis per active site.

The infinitesimal contribution to the global TOF from a differential volume element dV is:

$$dTOF(\vec{r}, t) = \rho(\vec{r}) f(\vec{r}, t) k_{local(r,t)} dV \quad (2)$$

Integrating over the full catalytic volume yields the generalized expression:

$$\text{TOF}(t) = \iiint_V \rho(\vec{r}) f(\vec{r}, t) k_{local}(\vec{r}, t) dV \quad (3)$$

This equation allows direct incorporation of experimentally determined spatial fields (e.g., enzyme distribution, site accessibility, local reactivity). For cumulative catalytic output, we integrate overtime to obtain the generalized turnover number (TON):

$$\text{TON}(T) = \int_0^T \text{TOF}(t) \, dt = \int_0^T \left( \iiint_V \rho(\vec{r}) f(\vec{r}, t) k_{\text{local}}(\vec{r}, t) dV \right) dt \quad (4)$$

The following sections showcase representative catalytic systems modeled through the ACS framework, demonstrating how spatial and temporal descriptors regulate turnover dynamics in enzymatic, homogeneous, and heterogeneous catalysis.

## Transition State Theory (TST) as a Limiting Case of ACS Framework

While ACS describes catalytic activity as a three-dimensional integral over the spatial density of active sites, dynamic modulation, and local rate constants, TST emerges when the reaction is assumed to occur at a single dominant point in the catalytic space—the transition state. Under this assumption, only one spatial configuration, located at r*, significantly contributes to the overall turnover frequency. Moreover, the system is considered to be in a steady-state regime, with negligible temporal variations.

The general ACS expression then simplifies to a localized form, where the local rate constant at r* follows a Boltzmann-type distribution, given by:

$k = (k^B T / h) \cdot exp(-\Delta G‡ / RT)$

The density of active sites, the modulation factor, and the infinitesimal volume around the transition configuration can be grouped into a single term, *P‡*, which represents the effective population reaching the activated state. The resulting expression:

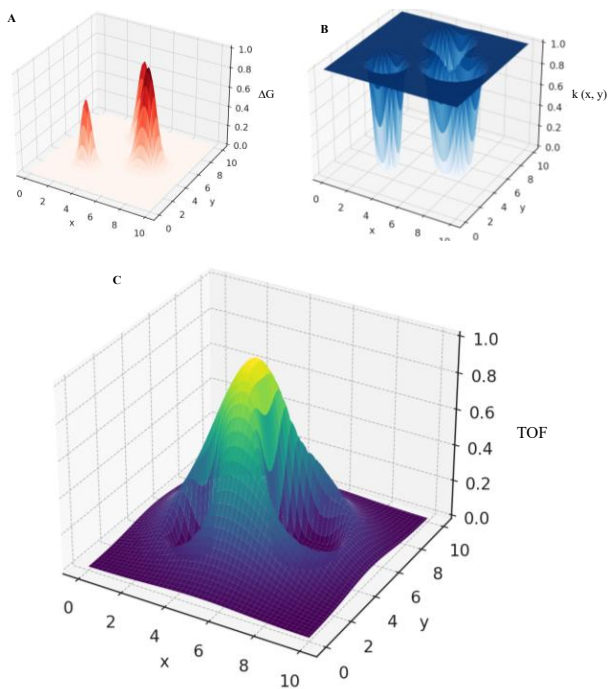

**Figure 2. Conceptual visualization of catalytic activity under the frameworks of TST and ACS.** (a) One-dimensional representation of an energy barrier $\Delta G\ddagger(x)$ and its corresponding local rate constant $k(x)$ according to TST, illustrating a single dominant transition state. (b) Two-dimensional energy landscape $\Delta G\ddagger(x, y)$ (left) and its corresponding local rate constant surface $k(x, y)$ from TST (right), highlighting multiple possible reaction pathways with spatially localized activation barriers. (c) Turnover frequency map $TOF(x, y)$ computed from the ACS model, integrating the spatial distribution of active sites $\rho(x, y)$, dynamic modulation $f(x, y)$, and local rate constants $k(x, y)$. This illustrates how ACS generalizes TST by accounting for multiple pathways, spatial heterogeneity, and dynamic accessibility across the catalytic domain.

$$TOF = P\ddagger \cdot (k^B T / h) \cdot \exp(-\Delta G\ddagger / RT)$$

is formally equivalent to the classical TST rate equation.

This demonstrates that TST is a statistical approximation emerging from the continuous spatial description provided by ACS, valid when catalysis is dominated by a single configuration and the system approaches quasi-equilibrium behavior. **Fig. 2** provides a comparative illustration of how catalytic activity is conceptualized under the TST and ACS framework. In **Fig, 2A,** the classical TST is represented as a one-dimensional reaction coordinate in which the entire kinetic behavior is governed by a single, sharply defined energy barrier. The local rate constant, $k(x)$, follows an exponential Boltzmann distribution and reaches its maximum precisely at the location of the transition state. This view inherently assumes spatial homogeneity and neglects the possibility of multiple concurrent pathways or spatial modulation. **Fig, 2B** extends this idea into a two-dimensional catalytic landscape, revealing multiple activation regions $(\Delta G\ddagger(x, y))$ and their corresponding local rate constants. While TST can in principle account for multiple transition states, it still treats each one independently and statically, lacking a framework to describe how these regions compete or cooperate dynamically.

**Fig. 2C** demonstrates how the ACS model overcomes these limitations by incorporating the spatial distribution of catalytic sites ($\rho(x, y)$) and dynamic factors such as accessibility or occupancy ($f(x, y)$). The resulting turnover frequency map ($TOF(x, y)$) emerges from the convolution of structural, energetic, and dynamic parameters. Unlike TST, ACS does not reduce catalysis to a single configuration, but instead captures the full heterogeneity of the system, including scenarios where multiple pathways are active simultaneously or when reaction dynamics are influenced by local conditions. This comparison highlights that TST emerges as a special case of ACS under the limiting assumptions of spatial localization, kinetic uniformity, and static modulation. The ACS framework thus provides a more general and physically realistic description of catalysis, particularly relevant for complex systems such as enzymes, heterogeneous surfaces, or dynamic interfaces.

## Application for Heterogenous Catalysis

We employed the ACS to analyze the hydrogen evolution reaction occurring at a single platinum atom. The starting point is the general equation for TOF, expressed as a volume integral. For a single atom of platinum, it is reasonable to assume a highly localized density of sites. We approximate $\rho(\vec{r})$ as a constant value within a small spherical volume V surrounding the atom, normalized such that:

$$\int_V \rho(\vec{r}) \, dV = 1 \tag{5}$$

$$\rho(\vec{r}) = \frac{1}{V} \tag{6}$$

$$V = \frac{4}{3}\pi R^3 \tag{7}$$

V is the volume of the sphere of radius R. Substituting this into the TOF equation 3 and assuming constant local kinetics:

$$TOF(t) = \iiint_V \frac{1}{V} \cdot 1 \cdot k_0 \, dV \tag{8}$$

Thus, the TOF reduces to:

$$TOF(t) = k_0 \tag{9}$$

This model demonstrates that the TOF is directly determined by the local reaction rate constant for this single-site system. It also calls attention to the utility of breaking down the integral into well-defined terms—site density, modulation function, and kinetic constant—each of which can be adjusted or refined in future studies to account for more complex behavior. In this case, our analysis shows that even a single platinum atom can be effectively described using a dimensionally consistent integral framework.

In order to test the consistency of the ACS theory in a more complex system, we consider a platinum catalyst with two exposed crystalline planes, {111} and {100}, and use known kinetic values from the literature. For the {111} plane, site density: $\rho_{111} = 1.5 \times 10^{15}$ sites/m$^2$ (a typical value for the tightly packed {111} plane); local reaction rate constant: $k_{loc al,111} = 0.1$ s$^{-1}$; and exposed area: $A_{111} = 0.5$ m$^2$. For the {100} plane, site density: $\rho_{100} = 1.2 \times 10^{15}$ sites/m$^2$ (less densely packed than the {111} plane), the local reaction rate constant, $k_{loc al,100} = 0.05$ s$^{-1}$.

According to the ACS framework, the TOF is the sum of the contributions from each plane:

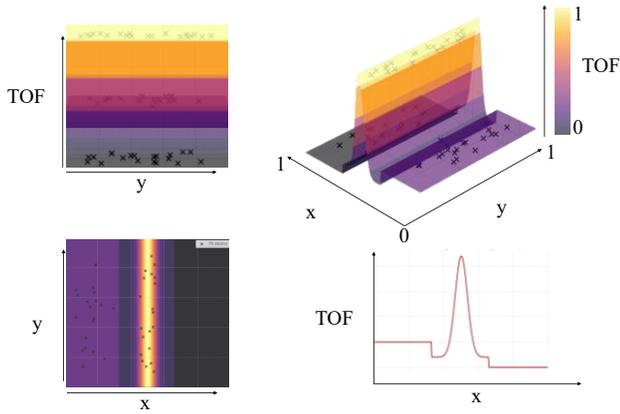

**Figure 3. Three-dimensional representation of the ACS for a platinum catalyst featuring {111}, {100}, and edge sites.** The colored surface indicates the local TOF, with a peak in the central region representing highly active edge sites. The transparent surface reveals the spatial distribution of platinum atoms (black spheres) across the catalyst. Higher TOF values are observed on the {111} facet (left), lower values on the {100} facet (right), and maximal activity at the edge zone (center). The surface was generated by assigning constant TOF values to the {111} and {100} regions, and a Gaussian function was fitted across the central domain to simulate enhanced activity at the edge. The resulting TOF landscape captures both discrete facet contributions and a smooth transition through the edge, reflecting realistic spatial modulation of catalytic activity.

$$\text{TOF}(t) = \iiint_{V_{111}} \rho_{111}\, k_{\text{local},111}\, dV + \iiint_{V_{100}} \rho_{100}\, k_{\text{local},100}\, dV. \quad (10)$$

Since the site densities and rate constants are uniform across each plane, and the planes are assumed to have defined areas, this becomes:

$$\text{TOF}(t) = \rho_{111} k_{\text{local},111} A_{111} + \rho_{100} k_{\text{local},100} A_{100} \quad (11)$$

For the {111} plane:

$\rho_{111}$ = 1.5 × 10$^{15}$ sites m$^{-2}$,
$k_{\text{local},111}$ = 0.1 s$^{-1}$,
$A_{111}$ = 0.5 m$^2$.

Contribution from the {111} plane: TOF$_{111}$ = (1.5 × 10$^{15}$) (0.1) (0.5) = 7.5 × 10$^{13}$ reactions s$^{-1}$.

For the {100} plane:
$\rho_{100}$ = 1.2 × 10$^{15}$ sites m$^{-2}$,
$k_{\text{local},100}$ = 0.05 s$^{-1}$,
$A_{100}$ = 0.3 m$^2$.

Contribution from the {100} plane:

TOF$_{100}$ = (1.2 × 10$^{15}$)(0.05)(0.3) = 1.8 × 10$^{13}$ reactions s$^{-1}$.

Thus,

$$\text{TOF}(t) = \text{TOF}_{111} + \text{TOF}_{100} \quad (12)$$
= 7.5 × 10$^{13}$ + 1.8 × 10$^{13}$ = 9.3 × 10$^{13}$ reactions/s.

Literature[14-17] values indicate site densities of $\rho_{111}$ = 1.5 × 10$^{15}$ sites/m$^2$ and $\rho_{100}$ = 1.2 × 10$^{15}$ sites/m$^2$. The intrinsic reaction rate constants are assumed as $k_{\text{local},111}$ = 0.1 s$^{-1}$ and $k_{\text{local},100}$ = 0.05 s$^{-1}$, and the exposed areas as $A_{111}$ = 0.5 m$^2$ and $A_{100}$ = 0.3 m$^2$, respectively. The {111} region not only provides higher TOF per unit area but also dominates the total reactivity due to its larger exposed surface. This example underscores how ACS theory enables spatially resolved interpretation of catalytic contributions, capturing structural sensitivity, facet-dependent kinetics, and the resulting impact on macroscopic performance  In our model, the local TOF for each region was calculated by multiplying the site density by a facet-specific reaction rate constant: 0.1 s$^{-1}$ for Pt{111} and 0.05 s$^{-1}$ for Pt{100}. These values are consistent with experimental and theoretical data across various hydrogenation and electrocatalytic reactions. The resulting spatial map of TOF (**Fig. 3**) reveals a clear contrast between the two domains: the {111} region not only contributes more reactive events per unit area, but also dominates the overall catalytic output due to its larger exposed area (0.5 m$^2$ versus 0.3 m$^2$ for the {100} facet). ACS theory resolve spatially distributed contributions in real catalytic systems, enabling a unified quantification of structural sensitivity, facet selectivity, and their impact on macroscopic activity.

## Application for Enzyme Catalysis

Among all catalytic systems, enzymes represent a unique class defined by their structural complexity, high specificity, and often extraordinary reaction rates. Their activity is not only governed by well-organized active sites but also by dynamic conformational changes and microenvironmental modulation, making them ideal candidates for testing the limits and versatility of the ACS framework. For our enzyme catalyst, we selected bilirubin oxidase (BOD). BOD is a multicopper oxidase that catalyzes the four-electron reduction of molecular oxygen to water, a process critical in both biological systems and practical applications such as biofuel cells.[11] Its selection for the ACS framework stems from its clearly defined active sites (a mononuclear T1 copper site for electron transfer and a trinuclear TNC cluster for oxygen reduction) and its well-documented catalytic behavior, making it an excellent model for illustrating how spatial and temporal descriptors influence enzymatic turnover dynamics. By studying BOD within the ACS framework, we can better understand how enzymatic architecture and environmental modulation interact to drive highly efficient catalytic processes. The integral volume can be reduced to a discrete summation over the individual catalytic sites due to the enzyme's well-defined structure. Each copper center in the enzyme corresponds to a distinct region of catalytic activity. This leads to a simplified, discrete version of the ACS expression:

$$\text{TOF} = \iiint_V \rho f\, k_{\text{local}}\, dV \quad (13)$$

$$k_{\text{local, T1}} = k_0 \exp\left[-\frac{\lambda}{k_B T}\left(1 - \frac{\Delta G}{\lambda}\right)^2\right] \quad (14)$$

$$k_{\text{local, TNC}} = \frac{k_{\text{cat}}[O_2]}{K_m + [O_2]} \quad (15)$$

$$\text{TOF} = \iiint_{V_{\text{T1}}} \rho_{\text{T1}} f_{\text{T1}} k_0 \exp\left[-\frac{\lambda}{k_B T}\left(1 - \frac{\Delta G}{\lambda}\right)^2\right] dV +$$

$$\iiint_{V_{\text{TNC}}} \rho_{\text{TNC}} f_{\text{TNC}} \frac{k_{\text{cat}}[O_2]}{K_m + [O_2]}\, dV \quad (16)$$

# Article

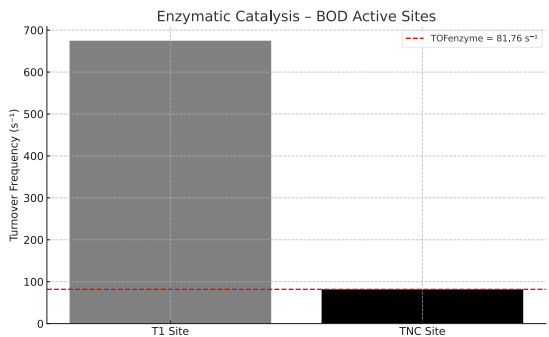

**Figure 3. –Resolved TOF in BOD.** The electron-transfer T1 copper site and the trinuclear cluster (TNC) responsible for oxygen reduction. Although the T1 site exhibits high intrinsic activity (TOF$_{t1}$ = 674.8 s$^{-1}$), the overall enzymatic rate is limited by the TNC site (TOF$_{tn}$c = 81.76 s$^{-1}$), as catalysis proceeds in series. The red dashed line indicates the rate-determining step, highlighting how local modulation constrains global catalytic efficiency in enzymatic systems.

The T1 copper site (which primarily performs electron transfer) and the TNC cluster (which catalyzes the reduction of oxygen to water). In the case of BOD, the T1 and TNC sites are coupled in series. The T1 copper accepts electrons from an external donor and transfers them to the TNC cluster. The TNC then uses these electrons to catalyze the reduction of O$_2$. Thus, while the TOF can be expressed as a sum over the individual sites' contributions, the enzyme's overall activity is constrained by the slower of the two processes. This introduces a dependency on the minimum rate of the two steps:

$$\text{TOF}_{\text{enzyme}} = \min(\rho_{T1} f_{T1} k_{T1}, \rho_{TNC} f_{TNC} k_{TNC}) \quad (17)$$

In this formulation, $\rho_{t1}$ = $\rho_{tn}$c = 1, since we are considering a single molecule of BOD where both sites are present, $f_{t1}$ and $f_{tn}$c are modulation factors reflecting conditions like local electron availability or oxygen concentration, $k_{t1}$ and $k_{tn}$c are the intrinsic rate constants for the electron transfer and oxygen reduction steps, respectively. For the T1 site, $f_{t1}$ = 1.0 (assuming optimal electron transfer conditions); $k_{t1}$ = 674.8 s$^{-1}$, a representative value from Marcu's theory modeling of electron transfer; and TOF$_{t1}$ = $f_{t1}$ $k_{t1}$ = 1.0 × 674.8 = 674.8 s$^{-1}$. For the TNC site, $f_{tn}$c = 0.85, reflecting slightly less than ideal modulation due to oxygen concentration or potential limitations; $k_{tn}$c = 96.18 s$^{-1}$, derived from Michaelis-Menten kinetics for oxygen reduction; and TOF$_{tn}$c = $f_{tn}$c $k_{tn}$c = 0.85 × 96.18 = 81.76 s$^{-1}$. Given that the two sites operate in series, the overall TOF is limited by the slower step, which is the TNC site in this case. Thus, the total TOF of the enzyme becomes:

$$\text{TOF}_{\text{enzyme}} = \min(\text{TOF}_{T1}, \text{TOF}_{TNC})$$
$$= \min(674.8, 81.76) = 81.76\, s^{-1} \quad (18)$$

Starting from the general ACS formulation, we discretize the TOF equation to account for the distinct catalytic sites within a single molecule of BOD (**Fig. 3**). By applying realistic rate constants and modulation factors, we determine that the overall enzymatic activity is limited by the oxygen reduction step at the TNC site, yielding a TOF of approximately 82 molecules of O$_2$ reduced per second. This approach demonstrates the ACS framework's flexibility in modeling molecular catalysts by directly incorporating site-specific kinetics and interactions.

## Application for Homogeneous Catalysis

To demonstrate the applicability of the ACS framework in homogeneous catalysis, we consider the prototypical case of a ruthenium polypyridyl complex, such as [Ru(tpy)(bpy)(OH$_2$)]$^{2+}$, used extensively as a water oxidation catalyst (WOC) in solution.[12,13] In such systems, the catalytic centers are molecular and fully solvated, but their activity is still influenced by local concentration gradients, conformational isomerism, and solvent organization, all of which are naturally embedded in the ACS formalism. We define catalytic volume V as a microscale region (e.g., within a nanodroplet or confined domain) containing the catalyst at a non-uniform spatial distribution due to aggregation, local coordination equilibria, or phase separation. The site density $\rho(\vec{r})$ is modeled as a Gaussian centered around a cluster core, representing a localized enrichment of the active species. The modulation function $f(\vec{r},t)$ incorporates temporal fluctuations in ligand geometry and redox state transitions, including Ru(II)/Ru(IV)/Ru(V) interconversions triggered by external voltage or light. The local catalytic rate $k_{local}(\vec{r},t)$ depends on the proton concentration and the local availability of water molecules, both of which vary with position due to microheterogeneity in the solvent environment.

$$\text{TOF}(t) = \iiint_V \rho(\vec{r}) \cdot f(\vec{r},t) \cdot k_{local}(\vec{r},t)\, dV$$
$$= 4\pi \int_0^{R_{max}} \rho(\vec{r}) \cdot f(t) \cdot k_{local}(\vec{r}) \cdot r^2\, dr \quad (19)$$

$$\rho(\vec{r}) = \exp\left(-\frac{r^2}{2\sigma^2}\right) \quad (20)$$

$$f(t) = 1 + 0.6 \cdot \sin(2\pi t) \quad (21)$$

$$k_{local}(\vec{r}) = 5 \cdot \exp\left(-\frac{r}{4}\right) \quad (22)$$

The integration of the ACS model over a Gaussian site density and exponentially decaying local rate profile yielded a mean turnover frequency of approximately TOF$_{mean}$ ≈ 723 s$^{-1}$, with oscillation boundaries defined by the dynamic modulation function. Specifically, the time-dependent fluctuation in catalytic accessibility resulted in a minimum turnover frequency of TOF$_{min}$ ≈ 289 s$^{-1}$ and a maximum of TOF$_{max}$ ≈ 1156 s$^{-1}$ (**Fig. 4**). These values are consistent with reported experimental ranges for homogeneous ruthenium-based water oxidation catalysts, further supporting the applicability of the ACS framework to solution-phase systems with moderate spatial heterogeneity and temporal gating behavior. pathways for spatially informed catalyst design in molecular systems.

## Toward Quantitative Implementation

The ACS framework provides a flexible and universal approach for analyzing catalysis across diverse domains. Its true power lies in our ability to experimentally measure or computationally predict its three core components: the spatial distribution of active sites $\rho(\vec{r})$ the dynamic modulation of site activity $f(\vec{r},t)$ and the local reaction rate $k_{local}(\vec{r},t)$ These components are not theoretical abstractions; they are accessible through well-established experimental techniques and are firmly rooted in decades of mechanistic research. We present a range of practical strategies (Table 1) that allow researchers to quantitatively characterize each of these functions in enzymatic, homogeneous, heterogeneous, photo-, and electrocatalytic systems, enabling deeper insight and more precise control over catalytic processes.

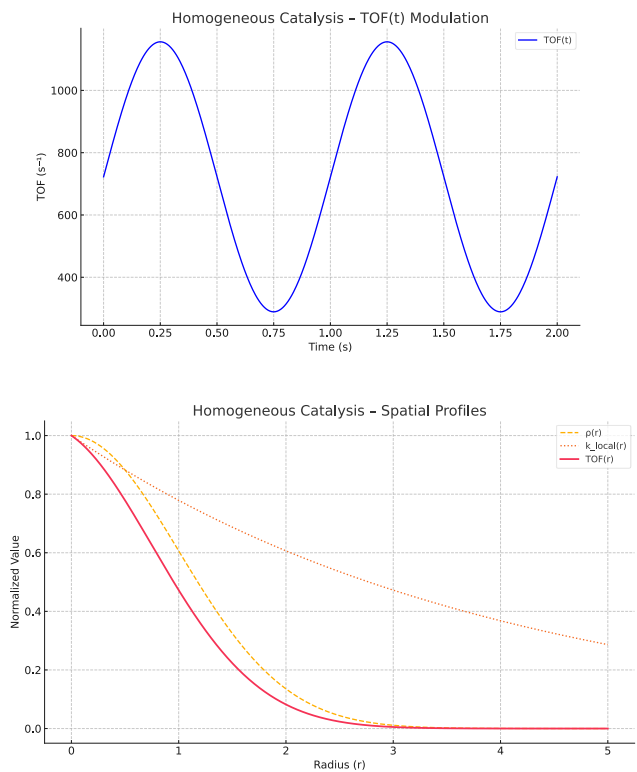

**Figure 4. Homogeneous Catalysis with Time-Dependent TOF Modulation.** A) Instantaneous TOF as a function of time for a homogeneous ruthenium-based catalyst undergoing temporal modulation via ligand fluctuations and redox transitions. The modulation function f(t) introduces dynamic changes in catalytic accessibility, yielding TOF oscillations between 289 s$^{-1}$ and 1156 s$^{-1}$. This dynamic behavior reflects the impact of temporal gating mechanisms in homogeneous systems under operando conditions. B) Spatial dependence of catalytic parameters in a homogeneous system. The site density ρ($\vec{r}$) follows a Gaussian distribution centered on a molecular cluster, while k$_{local}$($\vec{r}$) decays exponentially with radial distance. These spatial effects highlight the role of microheterogeneity and solvation in solution-phase catalysis.

For example, the spatial density of catalytically competent units can be accessed through imaging techniques (e.g., cryo-EM, SECM, AFM, fluorescence mapping), molecular dynamics, or known structural motifs. In enzymes, ρ($\vec{r}$) may reflect active pocket localization. In homogeneous catalysis, solvation and aggregation lead to spatial heterogeneity, which can be modeled using dynamic simulations or SAXS. In heterogeneous and electrocatalysis, ρ($\vec{r}$) reflects nanoparticle dispersion, defect distribution, and support structure. In photocatalytic systems, the spatial variation of ρ(r) can follow gradients of light absorption, photoactive phase concentration, or recombination-prone regions. Dynamic modulation (f($\vec{r}$,t)) function captures how site accessibility and activity fluctuate over time and space. In enzymatic catalysis, f($\vec{r}$, t) relates to conformational changes and allosteric transitions, often modeled via non-Markovian gating (e.g., renewal theory) or single-molecule trajectories. In homogeneous systems, solvent dynamics, redox cycling, or ligand rearrangements define f, and can be approached via time-resolved spectroscopy or kinetic modeling. For electrocatalysis, f($\vec{r}$,t) includes potential-dependent surface states, double-layer modulation, and ion accessibility, observable by impedance or operando SECM. In photocatalysis, f($\vec{r}$,t) captures the temporal availability of photogenerated carriers and field-driven separation under pulsed or modulated illumination, often accessible by transient absorption or intensity-modulated photocurrent spectroscopy.

In terms of local reaction rate (k$_{lo}$c$_{al}$($\vec{r}$,t)), the intrinsic reaction rate per site can be modeled using transition state theory (Eyring), Marcus–Hush–Chidsey formalism for charge transfer, or empirical relationships (e.g., Hammett, Brønsted). In enzymes, site-resolved rates can be modulated by pKa shifts or microenvironmental fields. In homogeneous catalysis, k$_{lo}$c$_{al}$ is tunable via ligand field effects and steric/electronic tuning and can be simulated by DFT or measured via stopped-flow techniques. In electrocatalysis, Tafel slopes and symmetrical factors help define local kinetics. For photocatalysts, k$_{lo}$c$_{al}$ depends on carrier lifetime, recombination probability, and charge injection dynamics, which can be deconvoluted using time-resolved photoluminescence, transient absorption, or surface photovoltage measurements.

Together, these approaches render the ACS framework operational across disciplines. Rather than replacing classical kinetic models, ACS embeds them within a broader spatiotemporal field description. In doing so, it preserves mechanistic detail while enabling cross-domain comparison and multi-scale interpretation. As spatially resolved techniques and dynamic probes become more prevalent, the ACS model offers a robust scaffold for understanding complex catalysis and guiding the rational design of next-generation systems. A unified theory of catalysis, by incorporating adjustable functionals that capture the unique characteristics of each system, would provide a common framework for directly comparing catalytic processes that are currently treated as distinct and unrelated. Rather than relying on separate descriptions for homogeneous, heterogeneous, and enzymatic catalysis, such a theory would employ a consistent set of parameters—site density, temporal modulation, and local rates—that can be tuned to match the specific properties of each catalytic system. This approach would enable meaningful comparisons, allowing researchers to quantitatively assess, for instance, how the efficiency of an active site on a solid catalyst compares to that of a molecular complex in solution, or how the stability and selectivity of a natural enzyme relate to those of an artificial catalyst.

The principal advantage of this framework lies in its ability to standardize the mathematical language and fundamental principles used to describe catalytic systems. Differences that previously seemed qualitative or difficult to measure would become quantitatively apparent. Instead of merely declaring one system "better" or "worse," researchers would gain precise insights into which factors—such as active site density, medium dynamics, or reaction barriers—are responsible for performance differences. This standardization would not only foster a deeper understanding of catalytic behavior across a diverse range of systems but also accelerate the development of new catalysts. By allowing direct comparisons, the best practices and insights derived from one type of catalysis could be systematically transferred to others, enabling more efficient innovation and a more cohesive view of the field as a whole.

# Article

**Table 1 - ACS Model –** Generalized Implementation Across All Types of Catalysis

| ACS Term | General Approach (All Catalysis Types) | Example Techniques |
| --- | --- | --- |
| **Site Density $\rho(\vec{r})$** | Derived from structural techniques (cryo-EM, AFM, SAXS, XRD, tomography) or inferred from catalyst design (nanoparticle dispersion, enzyme architecture, solvation environments). | Cryo-EM, SECM, AFM, super-resolution imaging, SAXS, operando XRD |
| **Dynamic Modulation $f(\vec{r},t)$** | Informed by operando dynamics: conformational fluctuations, redox state switching, adsorption/desorption cycles, or external stimuli (e.g., light, bias). Techniques include spectroscopy, transient kinetics, and impedance mapping. | Transient absorption, single-molecule FRET, impedance spectroscopy, operando SECM, time-resolved UV-Vis |
| **Local Rate $k_{local}(\vec{r},t)$** | Modeled via kinetic theory (Eyring, Marcus–Hush, Michaelis–Menten), or extracted from experimental data (Tafel slopes, rate constants, isotope effects, DFT/MD simulations). | Tafel analysis, Marcus theory, kinetic modeling, quantum calculations, isotopic labeling |

## Conclusion

This work introduces the concept of active catalytic space as a general, quantitative framework capable of describing catalysis in enzymatic, homogeneous, and heterogeneous systems alike. Through spatial integration of dynamically modulated activity fields, the ACS formalism not only captures known behaviors but also predicts turnover performance with high fidelity, as demonstrated across five distinct case studies. The approach transforms how catalytic activity is quantified, moving beyond rate constants and isolated mechanisms to a view where reactivity emerges from spatial orchestration and temporal evolution. As experimental techniques increasingly resolve catalytic fields in space and time, ACS stands as a robust scaffold for interpreting, comparing, and engineering catalysts in diverse chemical environments. This unifying theory opens new directions in catalyst design, data-driven discovery, and the fundamental understanding of reaction dynamics across scales.

## Methods

**Implementation of the ACS formalism**

The ACS model was implemented by numerically evaluating the generalized TOF as a spatially and temporally resolved functional over a defined catalytic domain. For all systems, the active region was discretized into finite elements (spherical or cylindrical coordinates depending on geometry), with mesh spacing adapted to the characteristic length scales of $\rho(\vec{r})$, $f(\vec{r},t)$, and $k_{local}(\vec{r},t)$. Gaussian, exponential, or empirical functions were assigned to each field based on either structural features (e.g., enzyme radius, electrode roughness) or literature-reported catalytic parameters. All volumetric integrals were evaluated using numerical quadrature (Simpson's rule or adaptive Gauss–Legendre routines, as appropriate), and time integrals were computed with second-order accuracy over time windows $\Delta t = 0.01–1$ s. Parameters for dynamic modulation functions $f(\vec{r},t)$ were either extracted from experimental kinetics (e.g., conformational gating frequencies) or modeled as sinusoidal or step functions to simulate external perturbation (light pulses, redox cycling, etc.). Field components were implemented in Python using NumPy and SciPy, and symbolic derivations were cross-checked with SymPy. For surface-integrated systems (e.g., single-atom electrocatalysts), the model domain was reduced to 2D with Dirac delta site distributions; for enzymatic and homogeneous systems, isotropic 3D volumes were used. The final TOF(t) values were benchmarked against known literature values for each case to ensure model consistency.

The spatial site density $\rho(\vec{r})$ was modeled as a Gaussian or Lorentzian profile centered at the active region, with standard deviation σ tuned to match structural resolution (e.g., 0.5–2.0 nm for enzymes, <1 nm for single-atom systems). Local catalytic rates $k_{local}(\vec{r},t)$ were assigned either as constants (for highly active centers) or as exponentially decaying functions of distance from the core. Modulation functions $f(\vec{r},t)$ were parameterized using characteristic timescales observed in cryo-EM dynamics, fluorescence correlation spectroscopy, or electrochemical impedance data, where available. To validate the ACS model, calculated TOF values were compared to experimental ranges reported in enzyme kinetics databases (BRENDA), electrocatalytic turnover studies, and photocatalysis under operando conditions.

**Mathematical Formalism of the ACS**

To rigorously describe catalysis as a spatially distributed and dynamically modulated phenomenon, the ACS framework formalizes TOF as a field quantity. Instead of considering catalysis as a localized, discrete event occurring at a single active site, ACS treats catalytic activity as a continuous distribution across a three-dimensional domain. This enables a general formulation in which local activity is integrated over space and time, capturing both instantaneous and cumulative behavior. The mathematical foundation of ACS is built upon three essential descriptors: the spatial density of catalytically active units, the time-dependent modulation of their accessibility or reactivity, and the intrinsic local reaction rate at each point. These three functions are combined to define an infinitesimal contribution to catalytic activity, which can be integrated to obtain global performance metrics.

By framing TOF and TON as integrals over spatial and temporal domains, the ACS formalism accommodates systems with complex geometries, dynamic behavior, and heterogeneous reactivity patterns. This includes biological catalysts with conformational gating, homogeneous systems with solution-phase fluctuations, and solid-state catalysts with spatially varying electronic environments. In the following sections, we present the theoretical structure of the ACS model and demonstrate its application to a range of catalytic systems. Additional complex systems are presented in SI, including enzymes such as carbonic anhydrase and lactate dehydrogenase, a prototypical single-atom platinum electrocatalyst, and a depth-modulated perovskite photoelectrode. Each case highlights how the ACS approach captures realistic catalytic behavior by integrating spatial architecture and dynamic modulation into a unified quantitative basis.

Numerical implementation involved discretizing the domain V into finite-volume elements in spherical, cylindrical, or Cartesian coordinates, depending on the system's geometry. The functions were parameterized using physically grounded analytical forms—e.g., Gaussian or Lorentzian, sinusoidal or sigmoidal waveforms for and exponential decay kernels or step functions. All integrals were computed using composite quadrature schemes (Gauss–Legendre and Simpson's rule), with convergence verified against analytical or semi-analytical benchmarks. Time evolution was resolved using second-order temporal schemes.

*Infinitesimal form of TOF in the ACS*

$dTOF(\vec{r},t) = \rho(\vec{r}) f(\vec{r},t) k_{local(r,t)} dV$

*Volumetric integral for generalized instantaneous TOF*

Integrating over the full catalytic volume yields the generalized expression:

$TOF(t) = \iiint_V \rho(\vec{r}) f(\vec{r},t) k_{local}(\vec{r},t) dV$

*Cumulative catalytic output as the general TON*

$TON(T) = \int_0^T TOF(t)\, dt = \int_0^T \left( \iiint_V \rho(\vec{r})\, f(\vec{r},t)\, k_{local}(\vec{r},t)\, dV \right) dt$

*Dimensional Analysis*

Dimensional analysis of the ACS integral ensures physical consistency for any catalytic system:

$TOF(t) = \iiint_V \rho(\vec{r}) \cdot f(\vec{r},t) \cdot k_{local}(\vec{r},t)\, dV$

$[TOF(t)] = \left(\frac{\text{sites}}{m^3}\right) \times (\text{dimensionless}) \times \left(\frac{\text{reactions}}{\text{site} \cdot s}\right) \times m^3 = \frac{\text{reactions}}{s}$

ACS provides a clear quantitative measure directly correlated to the overall catalytic activity, expressed as reaction rates in reactions per second. This confirms the internal consistency and applicability of the ACS framework across diverse catalytic systems.

**Acknowledgements**
This study was supported by the São Paulo Research Foundation (FAPESP), Brazil. The author thanks Professor Filipe C. Dalmatti A. Lima for reviewing the manuscript draft and providing valuable suggestions.

**Author contributions.** FNC developed the entire ACS theory.